# Human Genome Variation and the concept of Genotype Networks

Dall'Olio Giovanni Marco[1], Bertranpetit Jaume[1], Wagner Andreas[2,3,4], Laayouni Hafid[1]

[1] Institut de Biologia Evolutiva (CSIC-Universitat Pompeu Fabra), 08003 Barcelona, Catalonia, Spain

[2] Institute of Evolutionary Biology and Environmental Studies, University of Zurich.

[3] The Swiss Institute of Bioinformatics, Bioinformatics, Quartier Sorge, Batiment Genopode, 1015 Lausanne, Switzerland.

[4] The Santa Fe Institute, 1399 Hyde Park Road, Santa Fe, NM 87501, USA

## Abstract

Genotype networks are a method used in systems biology to study the "innovability" of a set of genotypes having the same phenotype. In the past they have been applied to determine the genetic heterogeneity, and stability to mutations, of systems such as metabolic networks and RNA folds. Recently, they have been the base for re-conciliating the two neutralist and selectionist schools on evolution.

Here, we adapted the concept of genotype networks to the study of population genetics data, applying them to the 1000 Genomes dataset. We used networks composed of short haplotypes of Single Nucleotide Variants (SNV), and defined phenotypes as the presence or absence of a haplotype in a human population. We used coalescent simulations to determine if the number of samples in the 1000 Genomes dataset is large enough to represent the genetic variation of real populations. The result is a scan of how properties related to the genetic heterogeneity and stability to mutations are distributed along the human genome. We found that genes involved in acquired immunity, such as some HLA and MHC genes, tend to have the most heterogeneous and connected networks; and we have also found that there is a small, but significant difference between networks of coding regions and those of non-coding regions, suggesting that coding regions are both richer in genotype diversity, and more stable to mutations. Together, the work presented here may constitute a starting point for applying genotype networks to study genome variation, as larger datasets of next-generation data will become available.

# Introduction

Genotype networks are a method used in the field of systems biology to study the "evolvability" or "innovability" of a set of genotypes having the same, broadly defined, phenotype (such as viability) and to determine whether a given phenotype is robust to mutations (Manrubia and Cuesta 2010; Wagner 2011). They have been used to study the evolvability of metabolic networks in simple organisms, by identifying how much a metabolic network can be altered without losing the ability of surviving using a given sugar as the only source of carbon (Wagner 2007; Matias Rodrigues and Wagner 2009; Wagner 2009; Samal et al. 2010; Dhar et al. 2011). Similarly, they have been used to study the ability of a metabolic network to "evolve" a new phenotype, such as the ability of surviving using other sources of carbon (Barve and Wagner 2013). Genotype networks have also been used to study how much RNA folds are "flexible", in terms of withstanding mutations without losing the secondary structure (Schultes and Bartel 2000; Ferrada and Wagner 2010).

The concept of genotype network is derived from the metaphor of "adaptive walks in the fitness landscape", proposed by Maynard Smith ((Maynard Smith 1970)1970), and has been expanded in successive literature (Orr 2005). John Maynard-Smith introduced the concept of protein space, a representation of all the possible protein sequences, as a framework to describe how evolutionary processes take place. This sequence space is "explored" by populations, which, mutation after mutation, and through generations of individuals carrying similar sequences, reach positions of optimal fitness. Genotype networks are derived from this concept, and in the literature they have also been referred to as neutral networks (Lipman and Wilbur 1991; Fontana and Schuster 1998). In this work, we prefer to use the term "genotype networks", because we do not have any information on the phenotype of the sequences (the individuals of the 1000 Genomes dataset are anonymous), and we do not know whether all the nodes in the network are effectively neutral to the fitness.

Genotype networks are also at the base of a model proposed to reconcile the two neutralist and selectionist schools of thoughts in evolution (Wagner 2008a). According to this model, evolution is characterized by cycles of "neutral" evolution, in which populations accumulate neutral, or even slightly deleterious, mutations, alternated by events of "beneficial" evolution, in which, after the appearance of a positive mutation, a new repertoire of genotypes accumulate in the population. In this case, genotype networks represent the set of genotypes in a population during a cycle of neutral evolution, and beneficial mutations are events that allow a population to switch from a genotype network to another. Under this model, even negative or neutral mutations can have a beneficial effect on the long run, as they allow the population to explore the genotype space and increase the chances of finding a beneficial mutation (Wagner 2008a).

In general, genotype networks are defined in relation to a given phenotype. For example, in the future they may be used to compare the variability of individuals with a genetic disease against a control dataset, or to study the genetic variation behind a phenotypic trait like lactose tolerance or eye color. However, in the current work we present only a background genome-wide scan of how the properties of the genotype networks are distributed along the genome, defining the phenotype as the "presence" of a genotype in any of the individuals of the 1000 Genomes dataset or in one of the populations. We executed coalescent simulations to predict whether the sample size of the 1000 Genomes is representative of the variation in real populations, and to verify how many samples are needed to represent networks of a given size. The genome-wide scan presented here is a base for future applications of genotype networks, and will allow to better implement this understand genome variation.

## Description of the genotype networks method

A genotype network is a graph whose nodes are genotypes such as DNA sequences (or, in our case, short haplotypes of Single Nucleotide Variants), and where two sequences are connected by an edge if they differ in a single nucleotide (Wagner 2011).

To better understand genotype networks, it is useful to first introduce the concept of Genotype Space, of which they represent a subnetwork. The genotype space is a network representing all the possible genotypes in a region of the genome: for example, Figure 1A shows the genotype space of a region of five contiguous Single Nucleotide Variants (SNVs). Each node in Figure 1A represents one possible genotype, as a string of "0"s and "1"s, where the "0"s represent the reference allele, and the "1"s represent the alternative allele. In this network, two nodes are directly connected if they differ only in a single nucleotide – e.g. the nodes "00000" and "00001" are connected.

In empirical data, populations usually do not occupy an entire genotype space, but only a small portion of it. As an example, in Figure 1B, we marked in green all the that are observed at least once in a hypothetical population. We define this portion of the genotype space as the "Genotype Network" of that population. The method we propose is based on studying the properties of these genotype networks in comparison with the whole genotype space, such as how much they are extended in the space, or whether the nodes are tightly connected.

In particular, we focus on two classes of attributes of genotype networks. The first class is suitable to understand how much of a genotype space is filled by a genotype network, and how the network is distributed throughout this space. The second class of attributes relate to the connectivity of the network, and has implications in the network's output when a point mutation appears.

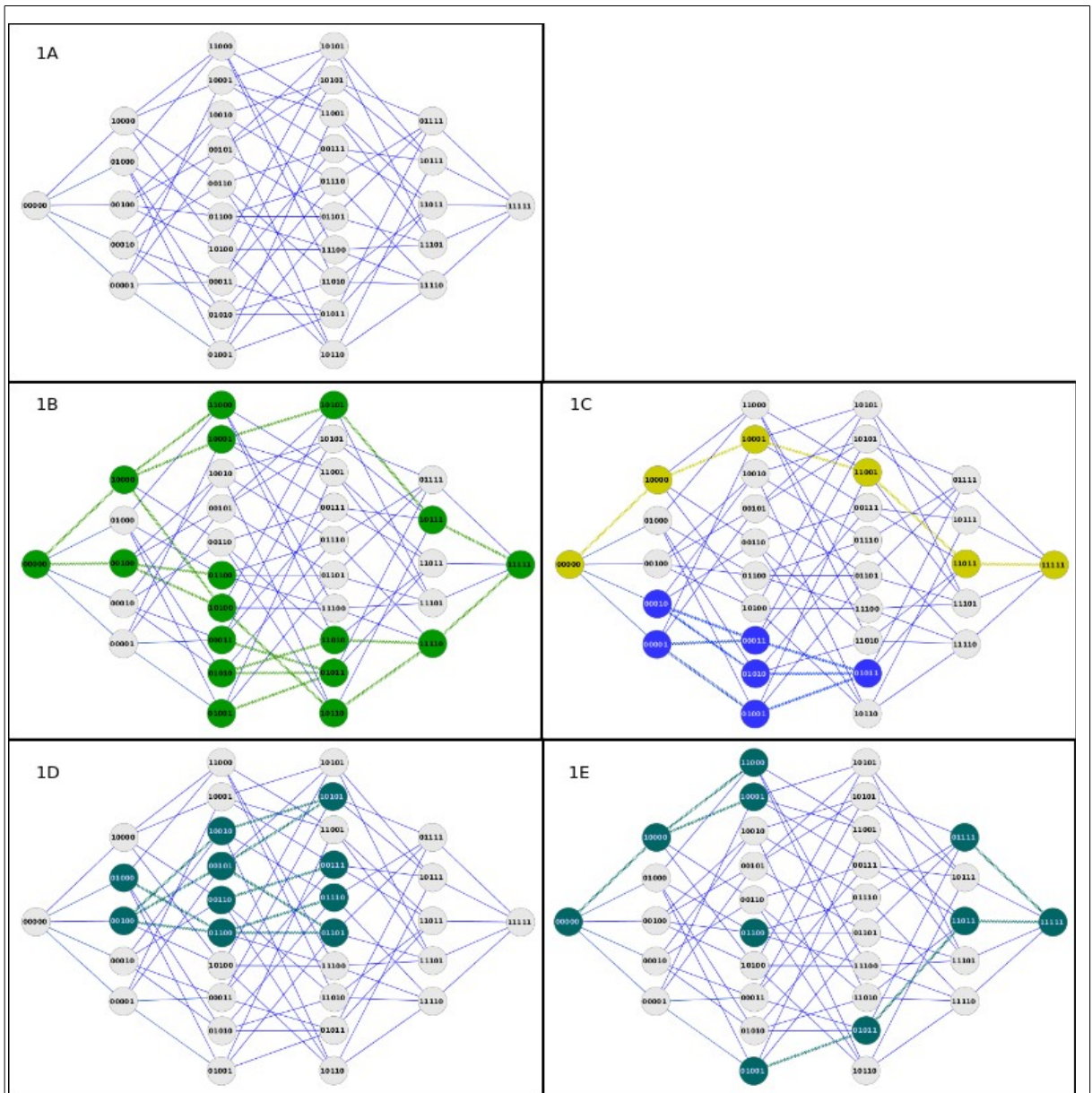

*Figure 1*: Examples of genotype networks and their properties. **A.** Representation of the Genotype Space for a region including 5 loci or Single Nucleotide Variants (SNVs). The space of all possible genotypes is represented as a Hamming graph (whole network). Each node represents one possible genotype, and each edge represents that the two nodes connected have only one difference. **B.** Example of genotype network. On top of the Genotype Space, we mark the genotypes observed in a population, and define it as the genotype network of that population (green nodes). **C.** Genotype networks of two populations (yellow and blue). The green population has a large average path length and diameter, while the blue population has a short average path length and diameter. **D.** Genotype network of a population having a high average degree and only one single component. **E.** Genotype network of a population having low average degree and many fragmented components.

**Extension and heterogeneity of the genotype network**

The first two attributes of interest are the number of vertices and the average path length of the genotype networks. These attributes allow to study how much the networks are extended in the space, and how much are the genotype in the population heterogenous, or genetically distant.

As an example of how two populations differing in their extension in the space and heterogeneity, Figure 1C shows the genotype networks of two hypothetical populations, one having a high average path length (yellow nodes), and the other a low average path length (blue nodes). Even though the two populations contain the same number of distinct genotypes (vertices), the yellow population is genetically more heterogeneous, and its individuals are more diverse genetically, than the individuals of the blue population. Specifically, the yellow population contains individuals that have very genetically distant genotypes, such as "00000" (all loci having the reference allele) and "11111" (all loci having the alternative allele), while in the blue populations, the genetically most distant individuals have at most two allelic differences (e.g. "00001" and "01001"). Thus, calculating the number of vertices and the average path length of the genotype networks of these two populations allow to identify that they are differently extended in the space.

**Robustness and stability of the genotype network**

Two other attributes of interest are the number of components and the average degree of the genotype networks. In other literature on genotype networks, these attributes have been interpreted as a measure of robustness to mutations (Matias Rodrigues and Wagner 2009; Payne et al. 2013). In this work, since we do not have a clear definition of the phenotype (we are using genotype networks based on the presence or absence of a genotype in a human population) we prefer to speak about stability of the genotype network of a genomic region to mutations.

Figures 1D and 1E show two hypothetical genotype networks that differ in the number of components and average degree. Both networks occupy the same number of nodes, but on average the nodes of the genotype network of Figure 1D (only the blue nodes) have more connections than the nodes in the network of Figure 1E. Specifically, in the network of Figure 1D, most nodes are connected to at least three other nodes, whereas the network in Figure 1E is much more fragmented, as most nodes have only one or two connections, and some groups of nodes are not even interconnected.

The importance of these two attributes in light of their biological significance can be explained by considering the effect of a random point mutation in a node of the network. If we randomly select a node and mutate one of its position, the resulting genotype will be one of the neighbor nodes in the network – this is because by definition, nodes in the genotype network are connected if they have only one mutation of difference. As the connectivity of the genotype network increases, there are higher chances that the mutant genotype will be a genotype that was already included in the network. For example, if we take the node '01100' in Figure 1D or 1E, and simulate a point mutation, the resulting genotype will be one of '01000', '00100', '11100'. '01110'. '01101'. In the case of Figure 1D, four out of these five genotypes already belong to the genotype network, so a point mutation will likely not alter the structure of the genotype network. In the case of Figure 1E, however, all the nodes connected to the original genotype do not belong to the genotype network, so any point mutation will be likely to change the structure of the network. Thus, we can interpret that genotype networks with a low number of connected components, and with high average degree, are more stable to mutations than the other type of genotype networks, as new mutations will be less likely to alter the structure of the network.

# Results

## Genome-wide distributions

We executed a genome-wide scan of genotype networks on the 1000 Genomes dataset, producing an overview of how the number of vertices, the average path length, the number of components, and the average degree of the genotype networks are distributed on the human genome. The scan is implemented in a sliding windows approach, dividing the genome into overlapping regions of 11 SNVs (see Methods). The results are available as a UCSC browser custom track hub, accessible at http://genome.ucsc.edu/cgi-bin/hgTracks?db=hg19&hubUrl=http://bioevo.upf.edu/~gdallolio/genotype_space/hub.txt . Raw data can be downloaded using the UCSC Tables function or forwarded under request.

Table 1 presents an overview of the regions having higher values in the genome, for each of the network properties calculated. Interestingly, most of these top regions belong to genes related to acquired immunity, such as HLA and MHC genes. In particular, all the three regions with higher average path length belong to HLA genes (HLA-DRB1, HLA-DRB5 and HLA-DQA1), while a region in HLA-DPA1 shows an exceptionally number of connected components. Moreover, if we divide the number of vertices by the number of components together, we find that two regions related to the MHC I complex, MICA and MICB, have very large components.

The higher genetic heterogeneity (in terms of the average path length and larger component size) of these regions involved in acquired immunity can be explained by their role in interacting with diseases. The HLA and MHC regions are already known for being among the most variable regions within human populations, and their function in interacting with pathogens greatly increases the genetic variability between individuals (Noble and Erlich 2012; Cao et al. 2013). The higher genetic heterogeneity of many regions involved in acquired immunity can be interpreted as a high capacity for finding "evolutionary innovations", probably in the form of response to different classes of pathogens, or to differentiation of the immune system. In this sense, the exceptional high number of components in the HLA-DPA1 region seems to be a contrasting result, as so many components indicate a very fragmented network. However, it may also be the case that the diversity of this region is so high that our sample size is not able to capture it, thus identifying a fragmented network instead of a large connected component.

**Evaluating the effect of missing samples**

A difficulty in applying genotype networks to SNV data is that, in order to correctly reconstruct the genetic variation in real populations, we need a very large number of samples. If the number of samples in the dataset is not enough, some nodes may not be represented in the network, not because they are not present in the real population, but just because they are missing in the samples. In particular, some properties such as the average path length and average degree can not be calculated properly (in mathematical terms) when there are too many missing nodes, so it is important to understand the effect of sample size on our ability to reconstruct the genotype networks.

To evaluate the effect of missing samples, we used a dataset of coalescent simulations including 5,000 haplotypes (2,500 diploid individuals) for each of the African, Asian and European populations, for a total of 15,000 haplotypes. From this dataset, we successively sampled a number of randomly chosen haplotypes, up to the point of having only 100 haplotypes per population (50 individuals), and we observed how the properties of the genotype networks varied as the sample size was reduced. The results are shown in Figure 2; each point in the figure represent the average of 5 independent re-sampling of the same number of individuals, using networks of 11 SNVs.

Figure 2 shows that, from a sample size of 5,000 to 1,000 haplotypes, the properties of the genotype networks have an almost linear distribution, suggesting that the effect of missing samples is not excessively strong. Interestingly, the relative differences between populations remain the same independently of the sample size: e.g., european and asian populations have always a similar distribution, and the global population has always a higher number of vertices, average path length, and average degree than the three sub-populations. For lower sample sizes, from 1000 to 100 haploypes (corresponding to 500 to 50 diploid individuals), the distribution of the properties of the networks start to decline, suggesting that the effect of missing samples is significant and may cause a wrong interpretation of the results. For example, for a sample size of 150 haplotypes, the global population has a number of components higher than the african population, reversing the result observed for higher sample sizes. Overall, this analysis show that, for networks of 11 SNVs of size, only the results based on more than 1,000 samples should be trusted.

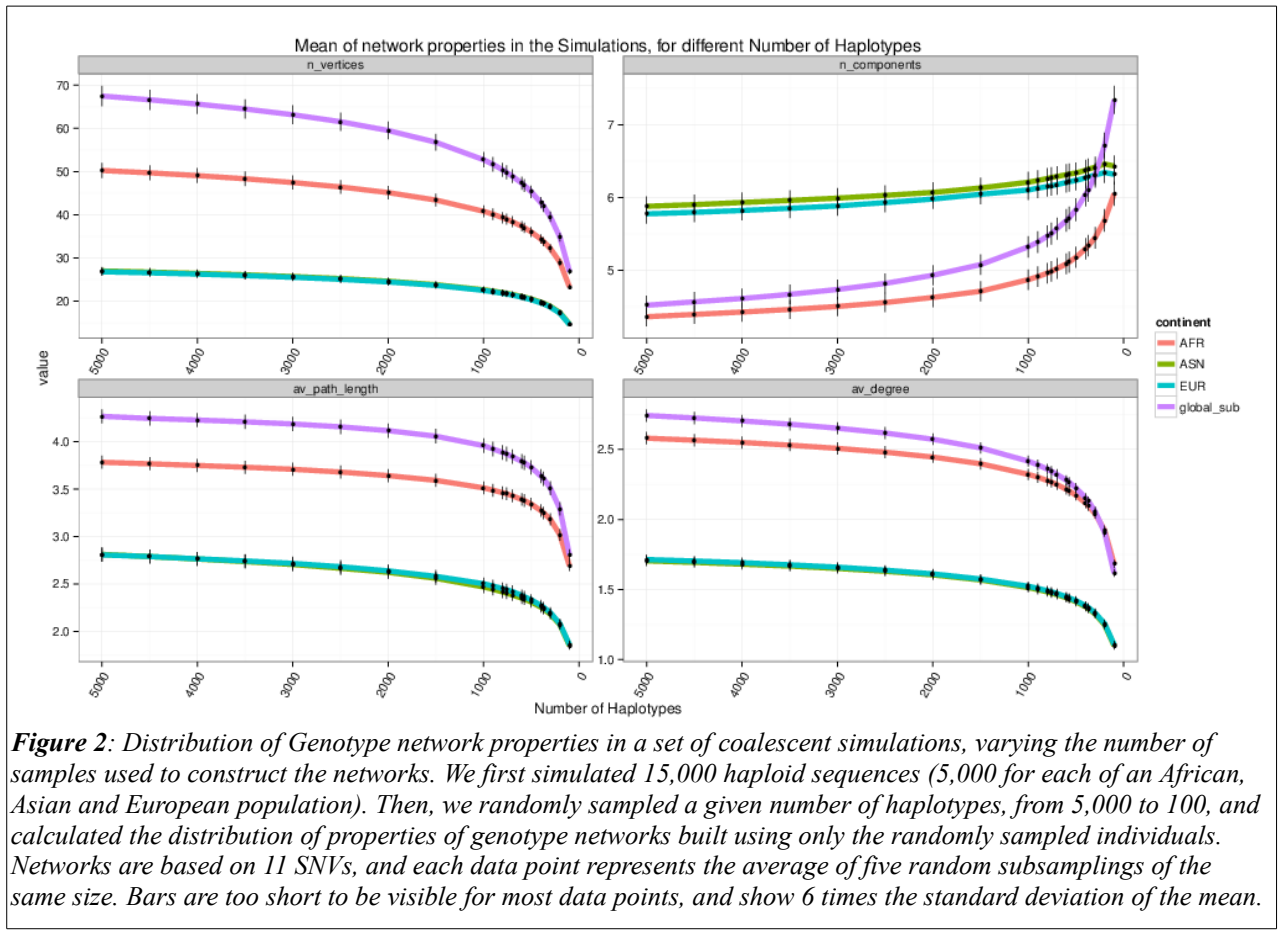

*Figure 2*: Distribution of Genotype network properties in a set of coalescent simulations, varying the number of samples used to construct the networks. We first simulated 15,000 haploid sequences (5,000 for each of an African, Asian and European population). Then, we randomly sampled a given number of haplotypes, from 5,000 to 100, and calculated the distribution of properties of genotype networks built using only the randomly sampled individuals. Networks are based on 11 SNVs, and each data point represents the average of five random subsamplings of the same size. Bars are too short to be visible for most data points, and show 6 times the standard deviation of the mean.

**Correlation between Network Properties**

Figure 3 shows the pairwise correlations between the properties of the genotype networks for data based on chromosome 22. Each panel in the figure shows the pairwise distribution between two properties, one on the X axis, and other on the Y axis, as defined in the diagonal panels. For example, the left-bottom panel shows the pairwise comparison of the region size (on the X axis) and the average path length (on the Y axis), also indicating that there is a correlation with a Pearson coefficient of 0.089 between these two properties.

The first two rows and columns show the effect of region size and of recombination on the network properties of the genotype space. The region size is the length, in base pairs, of the region occupied by the network of 11 SNVs; in average, in the 1000 Genomes dataset, these windows correspond to regions of about 2-3kb. The average recombination is a measure of the recombination observed in the regions, comparing the first and the last SNV in the window of 11 SNVs, and is obtained from the 1000 Genomes website. Thus, these two properties allow to determine if the distribution of a given network property is influenced by the physical or genetic longitude of the region, and if we have to correct for these factors when comparing different sets of regions. Notably, there is a small but significant effect of recombination on the average degree (r=0.34), on the number of vertices (r=-0.29), and on the average path length (r=0.23). Given this relationship, we took into account recombination when comparing the distributions of network properties in some analyses (e.g. see the section "Genotype Networks of Coding and Non-Coding regions").

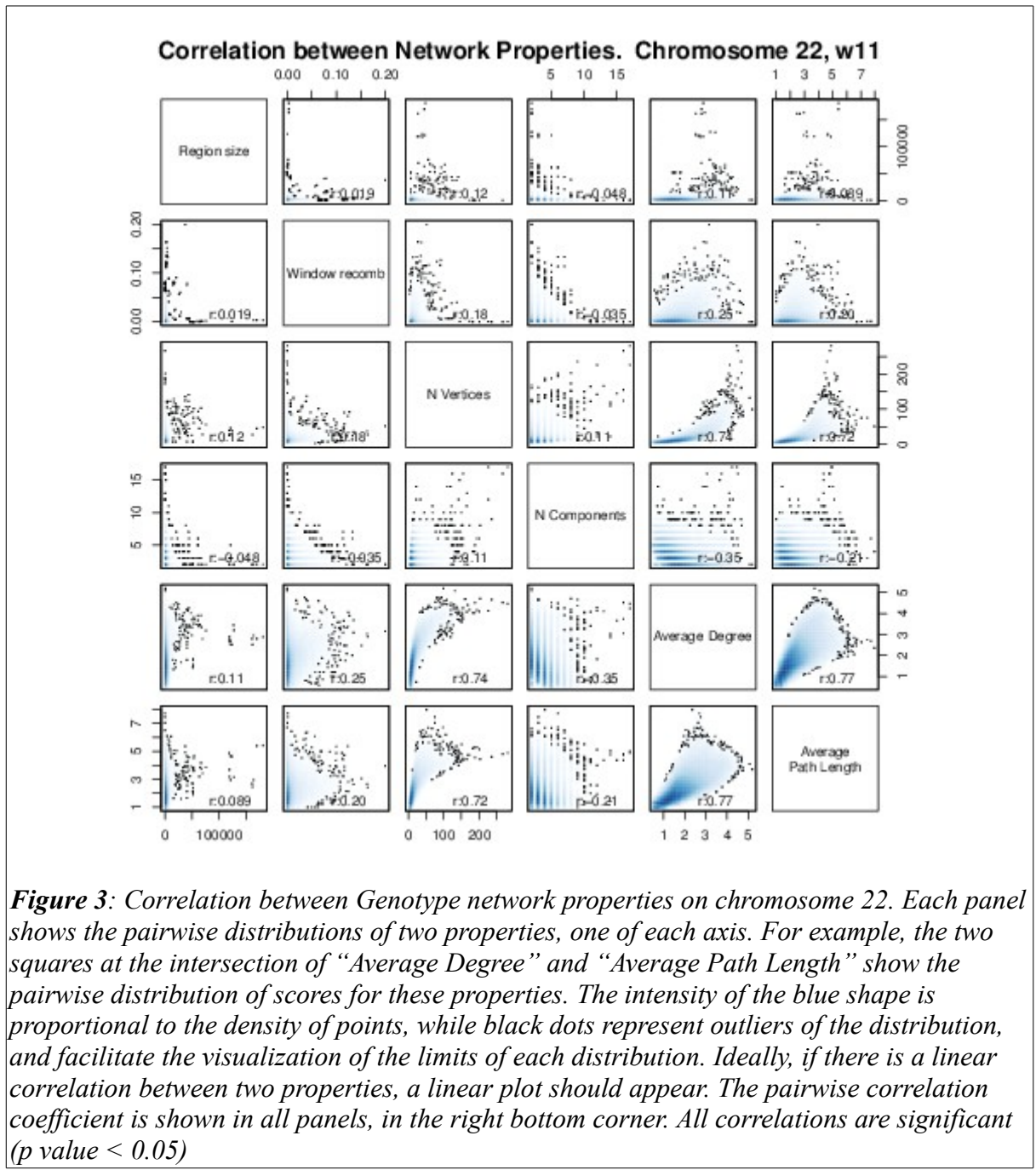

*Figure 3*: Correlation between Genotype network properties on chromosome 22. Each panel shows the pairwise distributions of two properties, one of each axis. For example, the two squares at the intersection of "Average Degree" and "Average Path Length" show the pairwise distribution of scores for these properties. The intensity of the blue shape is proportional to the density of points, while black dots represent outliers of the distribution, and facilitate the visualization of the limits of each distribution. Ideally, if there is a linear correlation between two properties, a linear plot should appear. The pairwise correlation coefficient is shown in all panels, in the right bottom corner. All correlations are significant (p value < 0.05)

The other panels in Figure 3 show the pairwise correlation between all the other network properties. Some properties of the genotype networks are correlated between each other. In particular, the average degree and the number of vertices have a correlation coefficient of 0.80, meaning that networks composed by more nodes tend also to have larger degree. This correlation can be due to the fact that in relatively small networks, a higher number of nodes in a network increases the probability that two nodes will be connected. Figure 3 also shows that the average degree and the average path length have a correlation coefficient of 0.70. This correlation may be also an artifact due that, for a network composed only by one or few nodes, the addition of one node to the network increases both the average path length and the average degree. However, as Figure 3 shows, for networks of larger size, the correlation between these two properties becomes weaker, giving a triangular/diverging pattern in the graph. Thus, as long as we do not analyze very small networks, we can consider the average path length and the average degree as independent variables.

**Genotype Networks of Coding and Non-Coding regions**

We used the functional annotations from the 1000 Genomes website to determine whether the presence of a coding or of a non-coding functional SNV has an effect on the properties of a genotype network. In particular, we classified all the networks into four categories, according to the functional effects of the SNVs included. The classes are:

I. networks containing only functional coding SNVs;

II. networks containing both functional coding and functional non-coding SNVs;

III. networks containing only functional non-coding SNVs;

IV. networks containing only SNVs for which no annotation is available and which have no known functional effect.

These annotations are based on Khurana et al. (Khurana et al. 2013). For simplicity, in the rest of the paper, we refer to these classes as "coding", "both", "noncoding", and "no annotations".

Since we have previously shown that the recombination rate is correlated with some properties of the network (see the section "Correlation between Network Properties"), we removed the networks having a high recombination rate (more than 1 cM between the first and the last SNV of the window). Moreover, to compare networks based on windows belonging to different annotation classes, we applied an analysis of covariance, using the annotation category as a grouping variable and the recombination rate as a covariable. This analysis aims at comparing the functional categories for genotype network properties after the recombination rate effect is excluded, and is performed separately for each continental group. Networks of the class "no annotations" (containing only SNVs for which no annotation is available) are excluded by this analysis, as a clear interpretation for this category cannot be provided.

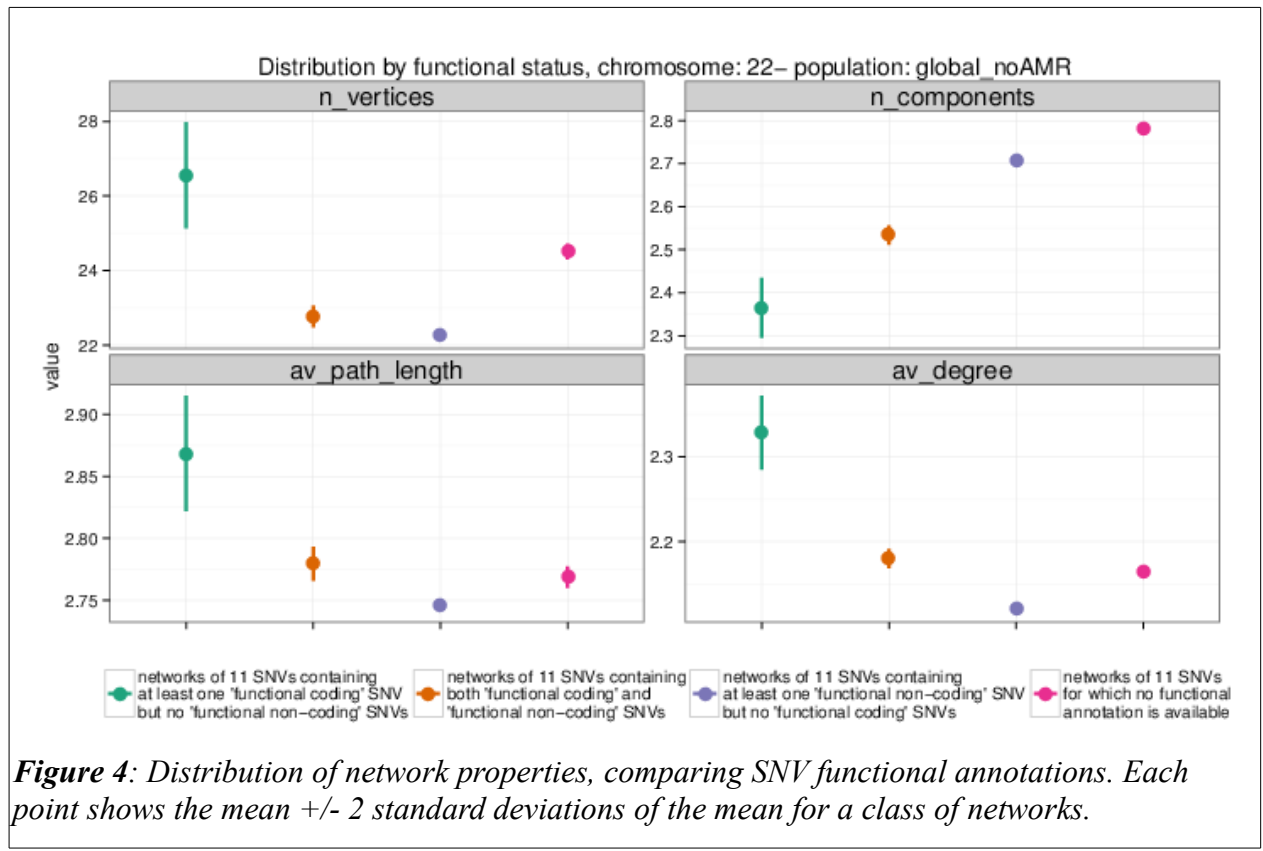

*Figure 4*: Distribution of network properties, comparing SNV functional annotations. Each point shows the mean +/- 2 standard deviations of the mean for a class of networks.

Overall, all three classes of networks have different number of connected components, a difference that is significant in all continental groups. Networks of class "coding" (containing only coding SNVs) have fewer connected components than the class "non-coding" (see Table 1). Moreover, networks of class "both" (containing both coding and non-coding SNVs) have intermediate values between the two other sets. The networks containing coding SNVs are thus more connected than the other classes, while non-coding networks tend to be slightly more fragmented. These differences hold in all cases when pairwise comparisons between the coding and noncoding classes are performed, even when a Bonferroni multiple testing correction is used ($p < 0.009$ in all comparisons).

Figure 4 shows the distribution of scores for chromosome 22 in the global population. Coding networks tend to have more vertices, greater average path length, and greater average degree, than non-coding networks. Overall, these results show that coding regions are less fragmented (they have fewer components) than non-coding networks, but at the same time, they are more extended in the genotype space (higher path length), and are more connected (higher average degree). It should be noted that even though the differences between annotation categories reach statistical significance in almost all comparisons and for almost all properties, the magnitude of the observed differences are small in general.

**Effects of a simulated selective sweep on the genotype networks**

Figure 5 compares the distribution of Network Properties between the data from chromosome 22 and two simulated datasets, representing a neutral and a selection scenario. The neutral scenario is based on the known demography for the European, Asian and African populations(Schaffner et al. 2005); the selection scenario is based on the same parameters using the neutral scenario, but adding a simulated sweep with a relatively low selection coefficient (0.015), in which the selected allele reaches a final frequency of 0.99. The choice of such selection scenario is justified because it has recently been proposed that strong selection events were rare in our evolutionary history (Alves et al. 2012; Messer and Petrov 2013), so a sweep with a low selection coefficient could be a more representative example of sweeps in the evolutionary history of humans.

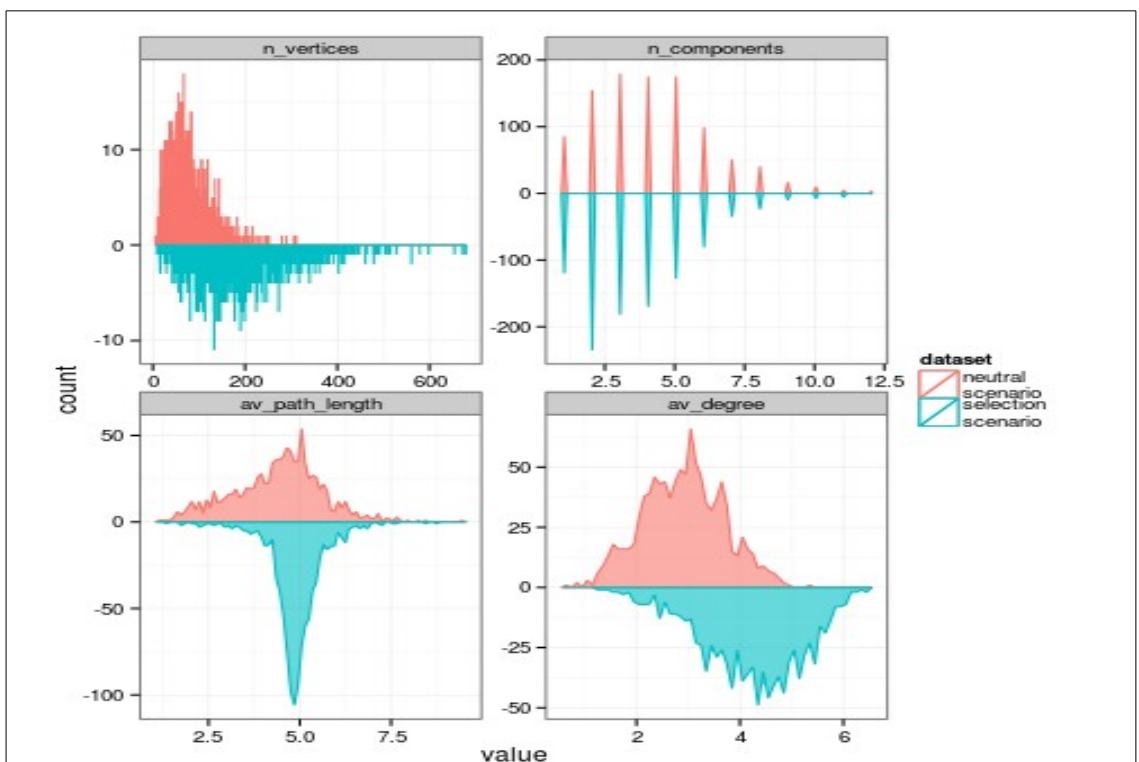

*Figure 5: Distribution of network properties, comparing a dataset of neutral demography simulations against a scenario of selective sweep. Selection scenario simulates a recent selective sweep with a selection coefficient of 0.015, and a final frequency of 0.99. The networks included in this graph are calculated by merging the 5,000 haplotypes of the three populations simulated (African + European + Asians) into a global population, and calculating the genotype networks on all the 15,000 haplotypes together.*

In general, the genotype networks of both neutral and selection scenarios differ in the distribution of all the four network properties considered. The less marked difference is in the number of components, for which the selection simulations have a slightly lower number of components than the neutral scenario (Wilcoxon test: W = 575321.5, p-value = 2.08e-09, alternative=two-sided; Kolmogorov-Smirnov test: D = 0.1152, p-value = 3.529e-06). For all the other properties the selection scenario leads to a higher number of vertices, average path length, and degree than the neutral scenario ($p<10E-15$ for both Kolmogorov-Smirnov and Wilcoxon test, for all the properties). In particular, the qqplots shown in Supplementary Figure 3 shows that the selection scenario has a higher proportion of values of average path length close to 4 than the neutral scenario. Together, these results indicate that, after a strong selective sweeps, the genotype networks tends to be both more stable and connected in the space (lower number of components and higher average degree), and at the same time more extended in the genotype space (higher number of vertices and average path length).

# Discussion

## Strategies to apply genotype networks to next generation sequencing data

So far, genotype networks have not been applied extensively to population genetics data. The main reason is that doing so requires very large datasets, on the order of thousands of sequences or more. Even in the work presented here, we limited our analysis to regions of 11 SNVs, because according to coalescent simulations, the number of samples in the 1000 Genomes dataset is only large enough to correctly reconstruct networks of this size. In the future, larger datasets will make it possible to analyze larger region; for the moment, the limitation of small sample size can be overcome by applying genotype networks in a sliding window approach, as presented in this paper. Thanks to this approach, it is possible to compare regions or genes of different size, by comparing the distributions of the properties of all the windows, instead of a single value per region.

Another difficulty in applying genotype networks to Single Nucleotide Variants data is that some properties of the network are slightly correlated with recombination. In particular, the degree and the number of vertices increase as the recombination rate in a region gets higher. Moreover, we can expect that a recombination event would fragment the genotype networks, creating networks divided into multiple unconnected components. In the implementation currently presented, there is no way to distinguish if the fragmentation of a network is due to recombination events, to population demography, or other factors. This difficulty can be partially solved by removing the windows that show higher recombination rates (likely to contain recombination hotspots), and by applying an analysis of covariance, using recombination as a covariable.

Another doubt is about whether Single Nucleotide Variants can really represent the genetic variability of the region, compared, for example, to genotype networks built on full sequences. In the present work, we used networks of 11 SNVs, which on average cover regions of about 2-3 kilobases in the genome. How much to the genotype networks built on SNVs can represent the full genotype space of these 2-3 kb? This question is difficult to answer; however, the type of variation captured by SNVs, which by definition are single position in the genome shown to be commonly variable in human individuals, should be enough to represent enough variation, if we limit our study to intra-species variation.

## Genotype Networks and human genome variation

In this work we presented a genome-wide scan of how the properties of the genotype networks are distributed on the human genome, using the data from the 1000 Genomes project.

One interesting result is that there is a small, but significant difference between networks of coding regions and those of non-coding SNVs. More specifically, networks including coding SNVs tend to have both a high average path length, and a high average degree, compared to the other group. This suggests that networks from regions including coding SNVs are both richer in genotype diversity (based on average path length), and at the same time are more stable to mutations (based on the average degree). These results can be combined with our analysis on simulated sequences, in which regions simulated under a selection scenario have also a lower number of components, but also a higher number of vertices, average path length, and average degree, compared to a region under neutrality.

Previous literature has shown that high connectivity and extension of genotype networks are associated to a greater "innovability", intended as the ability to find new phenotypes with different and evolutionarily novel functions (Wagner 2008b). Our finding that coding regions are more extended and connected is in agreement with this previous observation, as we can expect that evolutionary innovations are more likely to involve changes in coding rather than in non-coding regions. This hypothesis is further confirmed by the finding that some of the regions showing the most extended networks in the genome belong to genes involved in innate immunity, a function in which evolutionary innovations are important.

**Future Directions**

As the cost of genome sequencing will decrease in the next years, and as larger datasets of sequences will become available, genotype networks may become a useful tool to understand genome variation. An important possibility would be to apply them to datasets of case and controls individuals, to better understand the genetic variability behind a disease. In this case, genotype networks would be defined in reference to the presence or absence of a given disease, which is a more precise phenotype than the one used in this paper. Such analysis of the networks associated to disease may even allow us to learn how to identify variants in the genome associated with potential diseases. However, doing so will require very large datasets of case/control individuals.

In the present work, we showed how genotype networks can be applied to study intra-specific variation, in particular in the human genome. We provided an example of practices and approaches to use genotype networks for this type of variation. We showed that it is necessary to take into account the effect of recombination, and that some properties are correlated between themselves in empirical data. Moreover, we provided a description of the background distribution of these properties on the whole genome, and how they vary in coding and non-coding regions. Together, the work presented here may constitute a starting point for applying genotype networks to study genome variation, as larger datasets of next-generation data will become available.

## Methods

### Genotype Datasets and Individuals

We downloaded Single Nucleotide Variant (SNV) genotype data from the Phase I release of the 1000 Genomes dataset (ftp://ftp.1000genomes.ebi.ac.uk/vol1/ftp/phase1/analysis_results/integrated_call_sets/) (Durbin et al. 2010) on January 2013 (revision 2ff9d3af6cde in the repository, see "Reproducibility of the study"). Using the suite vcftools (Danecek et al. 2011), we removed all the SNVs having a minor allele frequency in the global population lower than 0.01, and a coverage lower than 2-fold. We considered only phased SNVs, and did not analyze chromosomes X and Y. A total of 11,684,193 SNVs passed this filtering, with an average of one SNV every ~250 bases.

From the 1000 Genomes dataset, we excluded all 242 American individuals (labels MXL, CLM, PUR, and ASW on the 1000 genomes website). One reason to exclude these populations is that it facilitates the comparison with the coalescent simulations, as no accurate demographic model for these populations is available (Schaffner et al. 2005). A second reason is that, based on a principal component analysis (not shown), these appeared to be genetically admixed with individuals from three other continents. The resulting dataset is composed of 850 individuals, or 1,700 haploid sets (chromosomes) grouped into individuals from three continents, African (AFR), Asian (ASN), and European (EUR). The african group includes 185 individuals (Yoruba from Nigeria and Luhyia from Kenya); the Asian group includes 286 individuals (Chinese from Beijing and South China, plus Japanese); the european group includes 379 individuals (Utah residents, Finland, Great Britain, Spain, Italy).

For the analysis of coding / non coding regions (see "Genotype Networks of Coding and Non-Coding regions) we used the functional annotations on SNVs from the 1000 Genomes ftp site (ftp://ftp.1000genomes.ebi.ac.uk/vol1/ftp/phase1/analysis_results/functional_annotation/annotated_vcfs). These annotations were generated by the 1000 genomes consortium, using the Variant Annotation Tool (Habegger et al. 2012; Khurana et al. 2013). The dataset of "functional coding" SNVs includes SNVs that are in a protein coding region, and that are transcribed and included in the mature transcript. The dataset of "functional non-coding" SNVs includes all the SNVs in non-coding regions that lie in transcription factor binding sites and in UTR regions, plus all the SNVs in regions that are transcribed but do not have any function, such as those in pseudogenes. All the other SNVs are included in a dataset called "no functional effect known", which includes all the SNVs for which no annotation is available. Intronic SNVs are included in this latter set, if there is no evidence for any functional effect.

## Construction of Genotype Networks

Genotype Networks are calculated using a customized version of VCF2Networks, a software produced by our group (https://bitbucket.org/dalloliogm/vcf2networks). VCF2Networks allows to parse a Variant Call Format (vcf) file (Danecek et al. 2011), generate a genotype network from it, and calculate network properties. The igraph library (Csardi and Nepusz 2006) and its python bindings are used to represent graphs and to calculate network properties.

Supplementary Figure 1 shows a scheme of the protocol used to convert a vcf file to a genotype network. The first step is to apply the Minor Allele Frequency filter of 0.01 described above, and to remove all SNVs that have unphased data, or are triallelic. Then, to generate the networks, we consider the two haplotypes of each individual as separate entities. Each genotype is encoded as a binary string, where "0" represents the reference allele, and "1" the alternative allele, using the annotations from the vcf files downloaded from 1000 Genomes (triallelic loci are not included in the 1000 Genomes dataset). After encoding all the distinct genotypes observed in a population, we build a network in which each node represents one genotype, and an edge connects two nodes if they differ in a single allele between each other, i.e., if the Hamming distance between their binary string representations is equal to one.

## Description of Network Properties

Among the properties implemented in the tools VCF2Networks, we calculated the following: the number of vertices, the average path length, the number of components, and the average degree. Here is a short description of how each of these properties are calculated.

The number of vertices is equivalent to the number of distinct genotypes present in a population. Notably, due to the definition of genotype network used here, the number of vertices is equivalent to the Dh statistics described by (Nei 1987). As an example, the network in Figure 1B has 17 vertices, while both networks in Figure 1C have exactly six vertices. The average path length is the average of all the possible shortest paths between pairs of genotypes in the network, and it corresponds to the average number of single nucleotide changes that it takes to move from any node in the network to another. In the example of Figure 1C, the yellow network has an average path length of 2.33, and the blue network has 1.67. Genotype networks of populations that have explored a greater portion of genotype space would have a higher number of vertices and a higher average path length.

A connected component of a graph is a subgraph in which each pair of nodes is connected through a continuous path. Thus, the number of components of a network is the number of subgraphs connected by at least one edge, and that are reachable without any "jump" between nodes. For example, the network in Figure 1B contains a single component, while the network in Figure 1E contains three connected components, as there are three groups of nodes that are not connected between each others. The degree of a node is the number of its neighbor nodes connected to it by an edge. For example, in Figure 1D, the node "01000" has a degree of one, as it is connected only to another node, while the node "01100" has a degree of four, as four edges emanate from it. The average degree of a network is the average of the degrees of all the nodes in the network: in Figure 1D, it is 2.20. Nodes without edges are called isolated and have degree zero. For networks with more than one component and some isolated nodes (e.g., Figure 1E), all components (including isolated nodes) are included in the calculation of the average degree. For example, the network in Figure 1E has an average degree of 1.54. As explained in the Introduction, we interpret the number of components and average degree as a measure of stability of the genotype network to point mutations.

**Sliding windows approach**

In order to apply genotype networks in a genome-wide scan, we divided the genome into contiguous and overlapping windows of 11 SNVs, building networks based on this fixed size. The choice of using 11 SNVs is justified after testing different window sizes on chromosome 22. More in detail, Supplementary Figure 2 shows how the properties of the genotype network of chromosome 22 vary, depending on the size used for the sliding windows approach. In particular, for a size of 11 SNVs, the networks of all the african, asian and european populations have a similar number of components, while, for larger sizes, these three populations start to diverge. Having the same number of components for all the population is important because in mathematical terms, calculating properties such as the average degree of networks with different number of components may lead to incomparable results. Another criteria to choose this window size is that the other properties of the network, such as the average path length and the average degree, show enough differentiation between the three populations.

**Calculation of Genome Wide top scores and filters**

To calculate which region showed higher values for each network property in the whole genome,

we first removed all the regions that were included in low quality regions, or in alignment gaps. To do so, we filtered out all the networks in which at least one SNV intersected one base with the "Gap" track in the UCSC Genome Browser (http://genome.ucsc.edu/cgi-bin/hgTrackUi?g=gap , last modification 2009-03-08). We also removed all regions corresponding to the centromeres, to the Giemsa bands neighbors to the centromers, and to the first and last Giemsa bands (for the telomeres) for each chromosome. Then, we applied a filter based on the quantile distribution of the remaining scores and on a manual inspection (to remove eventual artifacts) to select only the top scores for each network property.

**Simulations**

We implemented two sets of coalescent simulations, one based on the known demographic model (thus representing neutral evolution), and one simulating a selective sweep. We performed these simulations using the COSI software (Schaffner et al. 2005), version 1.2.1. Specifically, we simulated 3 populations (African, European, and Asian) of 5,000 individuals each, under the known demographic models for them (Schaffner et al. 2005). The parameters used for the simulations represent an out-of-Africa migration event 3,500 generations ago, followed by a split between European and Asian populations 2,000 generations ago, and, in the case of simulations with selection, a selective sweep in which the selected variant has a selection coefficient of 0.0150 and a final frequency of the selected allele of 0.99. The exact parameters used for the simulations are available in the repository of this project. After performing the simulations, we applied a filter of Minor Allele Frequency > 0.01, removing all the SNVs that had a low frequency in all three populations, i.e., we used the same criteria that we had used to filter the 1,000 Genomes data. These two datasets of simulations allowed us to estimate the distribution of network properties under a well-defined demographic model, and to estimate the distribution of these properties for a larger sample size (5,000 chromosomes per population). Moreover, these simulations allowed us to evaluate the effects of a strong selective sweep on the properties of Genotype Networks.

**Reproducibility of the study, and other tools used**

Following the best practices described in (Sandve et al. 2013) the whole project presented in this manuscript, including the raw data, the scripts to produce plots and analysis, and a versioned log of all the commands used, are available at https://bitbucket.org/dalloliogm/genotype_space.

Figure 1 was generated using the Cytoscape software (Smoot et al. 2011). To manipulate genome-wide data, we used the bedops (Neph et al. 2012) and the bedtools (Quinlan and Hall 2010) suites.

## Acknowledgements

This work was supported by grants BFU2010-19443 (subprogram BMC) awarded by Ministerio de Ciencia y Tecnología (Spain) and by the Direcció General de Recerca, Generalitat de Catalunya (Grup de Recerca Consolidat 2009 SGR 1101). GMD is supported by a FPI fellowship BES-2009-017731. We would like to thank Joshua Payne, Kathleen Sprouffske, José Aguilar-Rodríguez, and the members of the Andreas Wagner group in Zurich for useful feedback and help. We would like to thank Sergi Valverde and the members of the Evolutionary Systems Biology at the Institut de Biologia Evolutiva (IBE) in Barcelona, for discussion and support.

**Table 1:** regions showing top scores in the genome

| region | criteria | Closest gene | Distance to closest gene | Description of closest gene | 2nd closest gene | Description of 2nd closest gene |
|---|---|---|---|---|---|---|
| chr2:91959344-91968231 | high number of components. | GGT8P | inside gene | pseudogene | | |
| chr6:33037767-33038449 | high number of components. | HLA-DPA1 / HLA-DPB1 | inside gene | Homo sapiens major histocompatibility complex, class II | | |
| chr:7203189-7420319641 | high number of components. | ITGB8 | 50,684 bp | integrin | HLA-DPA1 / HLA-DPB1 | Homo sapiens major histocompatibility complex, class II |
| chr5:108634323-108635534 | high average degree. | PJA2 | 34,876 bp | praja ring finger 2, E3 ubiquitin protein ligase | AK021888 | unknown function |
| chr8:25935936-25937929 | high average degree. | EBF2 | inside gene | early B-cell factor 2 | | |
| chr6:32507854-32508257 | high average path length. | HLA-DRB1 | inside gene | Homo sapiens major histocompatibility complex, class II | | |
| chr6:32568909-32569343 | high average path length. | HLA-DRB5 | 11,297 bp | Homo sapiens major histocompatibility complex, class II | HLA-DQA1 | Homo sapiens major histocompatibility complex, class II |
| chr6:32611264-32611586 | high average path length. | HLA-DQA1 | inside gene | Homo sapiens major histocompatibility complex, class II | | |
| chr3:36921415-36921688 | high number of vertices. | TRANK1 | inside gene | tetratricopeptide repeat and ankyrin repeat Containing 1 | | |
| chr4:9176678-9178624 | high number of vertices. | C9JJH3 | 33,759 bp | Deubiquitinating enzyme | LOC650293 | transmembrane helix receptor |
| chr8:35105546-35106981 | high number of vertices. | UNC5D | inside gene | receptor of netrin involved in nervous system | | |
| chr4:9200148-9202368 | few components, but large number of vertices. | USP17L10 | 10,015 bp | Deubiquitinating enzyme | | |
| chr6:31357915-31358747 | few components, but large number of vertices. | MICA | 8,814 bp | MHC class I polypeptide-related sequence A | HLA-B | Homo sapiens major histocompatibility complex, class I |
| chr6:31455010-31456012 | few components, but large number of vertices. | MICB | 6,646 bp | MHC class I polypeptide-related Sequence B | uc003ntm.3 | HLA complex Group 26 (non-protein coding) |

**Supplementary Table 1**: Wilcoxon test comparing coding and non coding functional regions.

| Network Property | continent | Mean (coding Networks) | Mean (non-coding Networks) | Standard Deviation (coding Networks) | Standard Deviation (non-coding Networks) | Standard Deviation of the Mean (coding Networks) | Standard Deviation of the Mean (non-coding Networks) | statistic (Wilcoxon coding vs noncoding, Two-sided) | p.value (Wilcoxon coding vs noncoding, Two-sided) | p.value (Wilcoxon, Two-sided, Corrected Bonferroni) | is corrected p.value (Wilcoxon) Significant? |
|---|---|---|---|---|---|---|---|---|---|---|---|
| n_vertices | global | 26.5552 | 22.2909 | 23.4718 | 15.9832 | 0.7275 | 0.0473 | 64187032.5 | 1.05E-005 | 1.05E-003 | TRUE |
| n_vertices | AFR | 18.1595 | 15.5895 | 13.8822 | 9.7145 | 0.4303 | 0.0287 | 63976023 | 2.52E-005 | 2.52E-003 | TRUE |
| n_vertices | ASN | 11.3132 | 9.2384 | 10.7100 | 7.5662 | 0.3319 | 0.0224 | 66106856 | 4.90E-010 | 4.90E-008 | TRUE |
| n_vertices | EUR | 14.2305 | 11.8053 | 13.8575 | 9.2375 | 0.4295 | 0.0273 | 63651175 | 9.11E-005 | 9.11E-003 | TRUE |
| n_components | global | 2.3650 | 2.7084 | 1.1491 | 1.3327 | 0.0356 | 0.0039 | 50848306.5 | 7.60E-017 | 7.60E-015 | TRUE |
| n_components | AFR | 2.4409 | 2.8142 | 1.1166 | 1.3446 | 0.0346 | 0.0040 | 50354894 | 1.24E-018 | 1.24E-016 | TRUE |
| n_components | ASN | 2.4601 | 2.5714 | 1.1638 | 1.2175 | 0.0361 | 0.0036 | 56470300.5 | 3.56E-003 | 3.56E-001 | FALSE |
| n_components | EUR | 2.6628 | 2.8065 | 1.3332 | 1.2820 | 0.0413 | 0.0038 | 54909351 | 1.03E-005 | 1.03E-003 | TRUE |
| av_path_length | global | 2.8685 | 2.7457 | 0.7677 | 0.7864 | 0.0238 | 0.0023 | 65453759.5 | 2.30E-008 | 2.30E-006 | TRUE |
| av_path_length | AFR | 2.5356 | 2.3656 | 0.7465 | 0.7217 | 0.0231 | 0.0021 | 67443144 | 9.45E-014 | 9.45E-012 | TRUE |
| av_path_length | ASN | 1.8722 | 1.7307 | 0.7765 | 0.7282 | 0.0241 | 0.0022 | 66090994 | 5.79E-010 | 5.79E-008 | TRUE |
| av_path_length | EUR | 2.1187 | 1.9841 | 0.7844 | 0.7209 | 0.0243 | 0.0021 | 64921194 | 3.56E-007 | 3.56E-005 | TRUE |
| av_degree | global | 2.3283 | 2.1216 | 0.7195 | 0.6192 | 0.0223 | 0.0018 | 68435798.5 | 5.31E-017 | 5.31E-015 | TRUE |
| av_degree | AFR | 2.0392 | 1.8356 | 0.6364 | 0.5690 | 0.0197 | 0.0017 | 69679036.5 | 1.37E-021 | 1.37E-019 | TRUE |
| av_degree | ASN | 1.6062 | 1.4382 | 0.7564 | 0.6904 | 0.0234 | 0.0020 | 66928430.5 | 3.06E-012 | 3.06E-010 | TRUE |
| av_degree | EUR | 1.7444 | 1.5980 | 0.7503 | 0.6531 | 0.0233 | 0.0019 | 65007181.5 | 2.30E-007 | 2.30E-005 | TRUE |

***Supplementary Figure 1***: *Workflow used to calculate genotype network properties from a VCF file.*

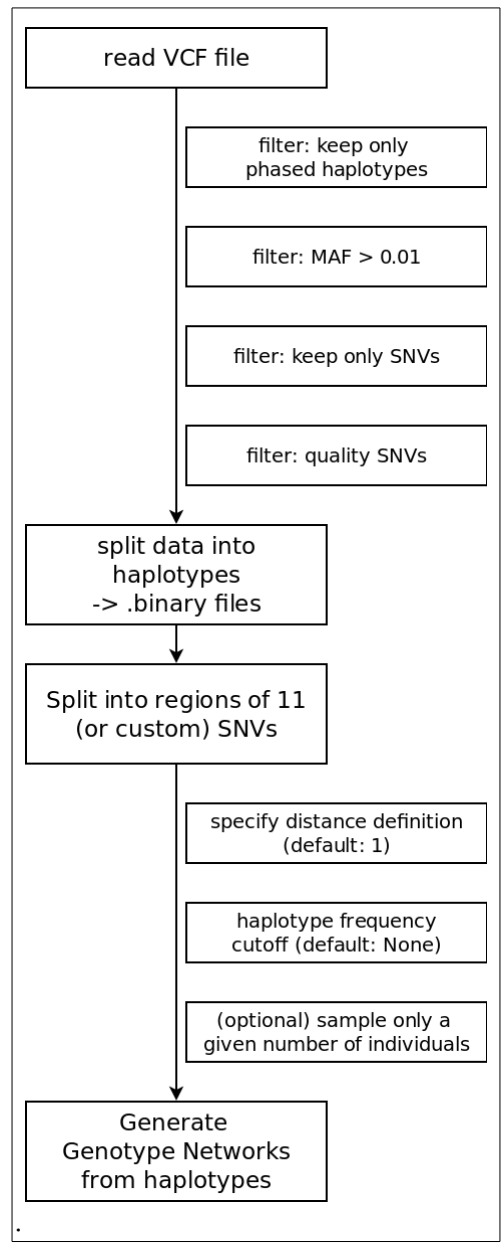

***Supplementary Figure 2***: *Distribution of genotype network properties in chromosome 22, changing the number of SNVs used to generate each network (window size), from 5 to 29 SNVs. In order to have the same number of individuals per each population, each point is based on 5 sampling of 370 haplotypes.*

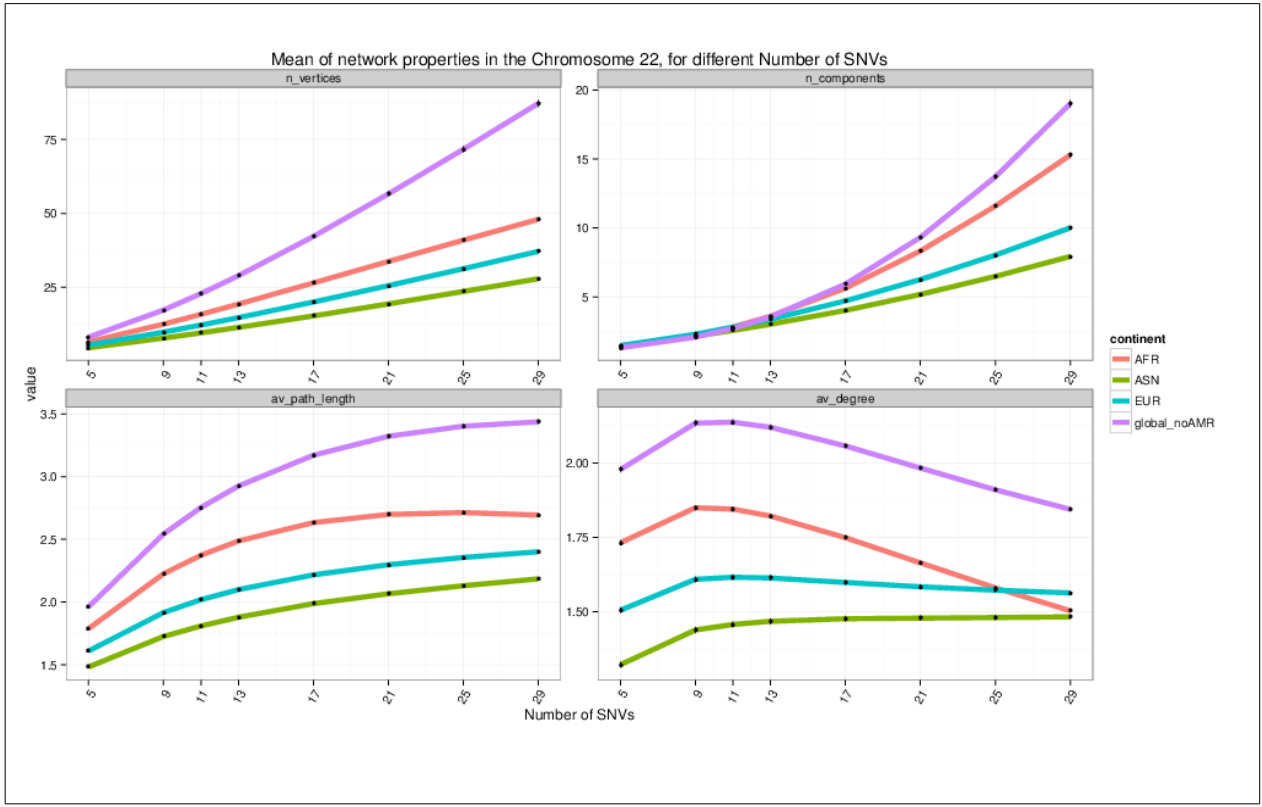

**Supplementary Figure 3:** *quantile-quantile plots (qqplots) of neutral vs selection simulations. Only the networks of the global populations (African + European + Asians) have been included.*

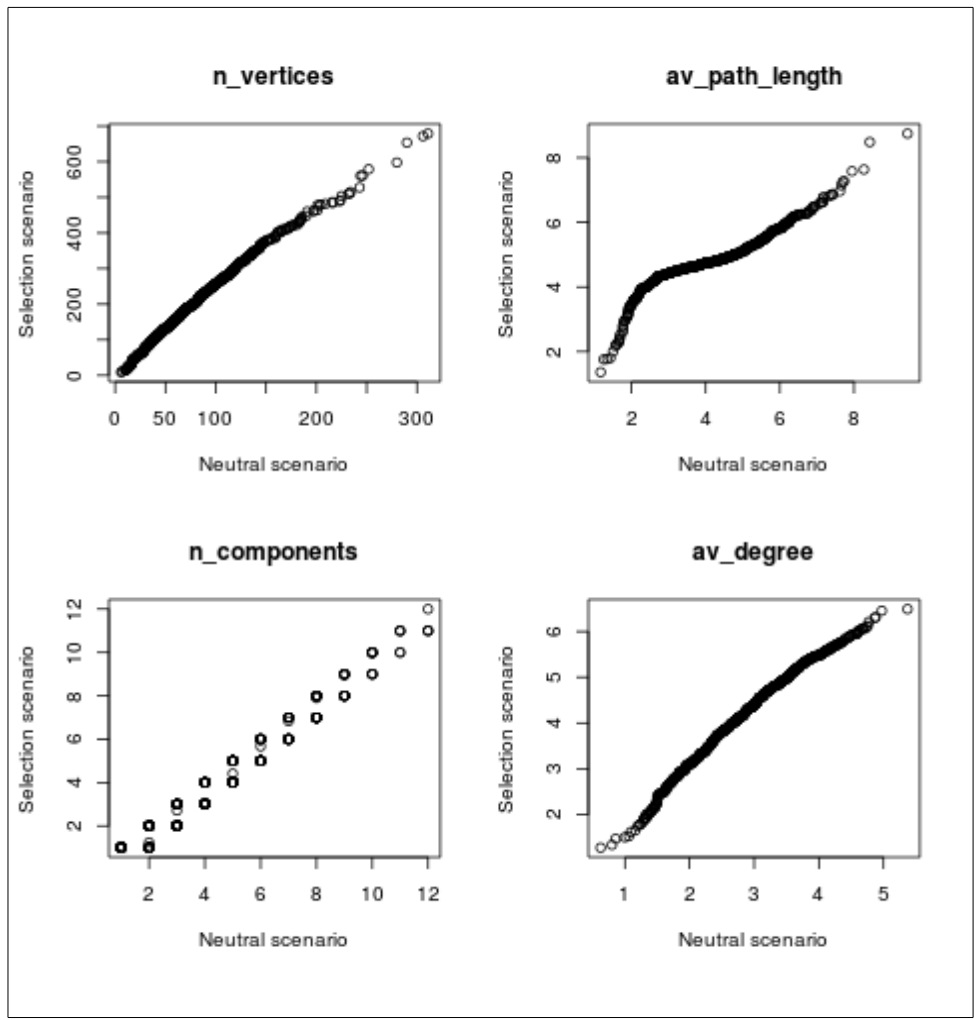